\documentclass[a4paper, 12pt]{article}
\usepackage[top=3.3cm, left=2.5cm, right=2.5cm, bottom=3cm]{geometry}
\usepackage{graphicx}

\usepackage[utf8x]{inputenc}
\usepackage[round]{natbib}
\usepackage{amssymb}
\usepackage{amsmath}

\newcommand{\x}{\mathbf{x}}

\newcommand{\R}{\mathbb{R}}
\newcommand{\E}{\mathbb{E}}
\newcommand{\rsigma}{\tau}

\usepackage[dvipsnames]{xcolor}

\usepackage{natbib}
\usepackage{multibib}
\newcites{App}{Appendix References}

\usepackage[pdfauthor={Tuomas A. Rajala}, hidelinks, colorlinks, citecolor=blue]{hyperref}
\usepackage{authblk}

\title{When do we have the power to detect biological interactions in spatial point patterns?}

\author[1]{T. Rajala\footnote{Corresponding author, \texttt{t.rajala@ucl.ac.uk}}}
\author[1]{S. Olhede}
\author[2]{D.J. Murrell}

\affil[1]{Department of Statistical Science, University College London, Gower Street, London WC1E 6BT, UK}
\affil[2]{Centre for Biodiversity and Environment Research, University College London, Gower Street, London WC1E 6BT, UK}

\begin{document}
\maketitle\footnotetext{Running title: "On power to detect spatial interactions"}


\section{Abstract}

\begin{enumerate}

\item Determining the importance of biotic interactions in assembling and maintaining species-rich communities remains a major challenge in ecology. In plant communities, interactions between individuals of different species are expected to generate positive or negative spatial interspecific correlations over short distances. Recent studies using individual-based point pattern datasets have concluded that (i) detectable interspecific interactions are generally rare, but (ii) are most common in communities with fewer species; and (iii) the most abundant species tend to have the highest frequency of interactions. However, there is currently no understanding of how the detection of spatial interactions may change with the abundances of each species, or the scale and intensity of interactions. Here, we ask if statistical power is sufficient to explain all three key results.

\item We use a simple 2-species model, where the scale and intensity of interactions are controlled to simulate point pattern data. In combination with an approximation to the variance of the spatial summary statistics that we sample, we investigate the power of current spatial point pattern methods to correctly reject the null model of pairwise species independence.

\item We show the power to detect interactions is positively related to both the abundances of the species tested, and the intensity and scale of interactions, but negatively related to imbalance in abundances. Differences in detection power in combination with the abundance distributions found in natural communities are sufficient to explain all the three key empirical results, even  if all species are interacting identically with all other species. Critically, many hundreds of individuals of both species may be required to detect even intense pairwise interactions, implying current abundance thresholds for including species in the analyses are too low.

\item Synthesis:  The widespread failure to reject the null model of spatial interspecific independence could be due to low power of the statistical tests rather than any key biological processes.  Our analyses are a first step in quantifying how much data is required to make strong statements about the role of biotic interactions in diverse plant communities, and power should be factored into analyses and considered when designing empirical studies.

\end{enumerate}

\paragraph{Keywords:} Determinants of plant community diversity and structure; 
Interspecific interactions; 
community ecology; neighborhood analysis; null model; spatial point patterns; statistical power

\section{Introduction}

Understanding the contribution of biological interactions to the assembly and regulation of  natural communities remains a key goal in ecology. The continual development and refinement of methods to detect interactions from spatial, temporal and spatio-temporal data has therefore been a mainstay of the literature on the subject. 

A particular focus on the role of competition can be found in plant ecology, not least because plants seem to require the same few 
nutrients~\citep{silvertown2004plant}, but also because their sessile nature might permit an understanding of processes from the spatial pattern of individuals~\citep{murrell2001uniting}, and allow for easier experimental manipulation ~\citep{goldberg1992patterns}.  Multiple methods exist to detect interspecific interactions but in non-manipulative field conditions there are often only two choices, both of which rely upon data where the location, identity and often size of every individual is recorded~\citep{Wiegand2017}. The first option is to fit growth and/or survival models that take into account the identity and size of nearby neighbours 
\citep[e.g.][]{uriarte2004spatially, Stoll2005, Comita2010, Stoll2015}. However, this requires repeated sampling over time in order to track the fate of every individual and very often such data is not available. Another issue is that because all interaction parameters are fitted at once, considering all pairwise interactions is very difficult due the large number of parameters. As a consequence neighbouring individuals are usually lumped into conspecifics and heterospecifics with the potential problem that interspecific interactions are 'lost' due to cancelling out of weak and strong, and/or positive and negative effects of different species.  The second option is to investigate the spatial pattern of the community to test the null hypothesis that species are independently arranged with respect to one another. Inference from a single snapshot of the community relies upon the assumption that spatial data retains the 'memory' of the birth, death and growth of the individuals 
\citep{Flugge2012} and consequently the effect of interspecific interactions should show up as inter-species spatial dependence after any effect of the abiotic environment has been removed \citep{murrell2001uniting}. Under the assumption that all pairwise tests are independent, each pair of species can be assessed individually, and dependent interactions are categorised as being a competitive interaction if the species are spatially segregated, and facilitation if they are aggregated in space, although confirmation via experimental manipulation is still advisable. Due to less restrictive data requirements (the test can be carried out on a single sampling of the community), the spatial snapshot option has proven to be very popular, and the test methods employ well-established spatial statistics such as Ripley's $K$ or the pair correlation function to test the null model of spatial independence~\citep{Wiegand2012b}.

The results of previous spatial analyses of multi-species communities have found only a very low frequency of interspecific spatial interactions (aggregation/segregation) over scales relevant to plant competition, implying interspecific interactions are generally rare, or weak (as discussed by \citealt{Luo2012,Wiegand2012b,Wang2014,Chacon-Labella2017}). However, comparisons of different plant communities have also revealed a positive relationship between the frequency of spatial independence and the number of species in the community \citep{Luo2012,Wiegand2012b,Wang2014,Chacon-Labella2017}. Spatial independence between all pairs of species is one of the key assumptions of null models for biodiversity \citep{McGill2010}, and the low frequency of detected interactions has been put forward in support of this assertion \citep{Wiegand2012b,Perry2014,Chacon-Labella2017}. However, classical niche theory also predicts the strength of interspecific interactions to decline as the number of coexisting species increases 
(equation 4 in \citealt{Chesson2000}), with the relative strength of interspecific interactions being proportional to $1/(s-1)$ for $s$ species. Therefore the main difference between the theories is that null models for biodiversity assume spatial independence for all communities regardless of species richness, whereas niche theory predicts spatial interactions are likely to be stronger, and therefore more frequently detected in less species-rich communities. Hence we argue the spatial analyses appear to better support the predictions of classical niche theory.

However, both the low frequency of interspecies interactions and the relationship between species richness and species interactions could arise due to the ability of the statistical tests to detect significant interactions at the sample sizes being used. Because of the unequal treatment of the null and alternative hypothesis in classical testing, failure to reject the hypothesis of no interaction does not provide concrete proof of a lack of interactions.
As pointed out by \cite{Wiegand2012b}, when species are rare the rate at which two species might co-occur in space is also very low and the statistical tests used might not be able to detect any interaction, even if it were very strong. If, as is often the case, species-rich communities have few common and many rare species, then we would expect to detect few significant interactions. Indeed, several investigations have found the frequency of significant spatial associations between species to be positively related to the abundance of both species being considered \citep{Luo2012,Wiegand2012b,Wang2014}, raising the possibility that interactions can only be detected amongst the most abundant species. 

For all tests a lower limit on the abundances of species to be included in the analyses must normally be set, and this acknowledges there is a limit to our ability to detect even strong interactions in small sample sizes. Previous investigations have used a range of lower abundance thresholds including 100 \citep{flugge2014method}, 70 \citep{Wiegand2012b}, 30 \citep{Perry2017} and even 18 \citep{Chacon-Labella2017} individuals. However, how and why is the lower threshold of individuals selected? What are the limits of our analyses to detect significant interspecific interactions? We are aware of no study that investigates the statistical power of the tests for spatial independence between pairs of species that are commonly used and consequently there are no guidelines for the lower abundance threshold. 
As such care is required when interpreting failures to reject the null hypothesis, and we argue it is hard to make strong statements about 
the relative roles of stochastic- and niche-based processes across different communities until we gain a better understanding of the power of the methods to detect departures from spatial independence. In other words, is spatial independence a good first approximation in species rich plant communities because of diffuse competition leading to weak interactions, or is it because the statistical methods lack the power to detect the interactions for the given sample sizes typically available?

Here we will elaborate on the statistical power of commonly used methods to detect significant interactions from spatial point pattern data. We shall study this problem by constructing a simple model for generating bivariate patterns where we can directly control the strength of interaction, and by utilising an approximation to the variance of the spatial summary statistic. We will show how the power to detect significant interactions is very much a function of the species' abundances, the strength of the interaction (normally the variable we are trying to infer, and therefore unknown), and the spatial scale over which the test is performed. Unfortunately, it is not possible to provide definitive sample size criteria since the power also changes with the summary statistic and test method being used. Despite this, we believe that even a rough understanding of the power of the tests to detect dependent structure is better than no understanding. With this caveat in mind, our analyses will suggest previous abundance thresholds for species inclusion are likely too low to detect even very strong interactions in the most species-rich communities being tested, thus questioning the previously derived conclusion of a lack of dependence between species. Since power can be estimated from Monte-Carlo simulations we hope our results will motivate ecologists to think more about the issue of sample size in future studies and therefore help to resolve the debate over the relative importance of biotic interactions in species-rich communities.

\section{Materials and Methods}

\subsection{Summary statistics for bivariate interaction}
Consider data for two species labelled 1 and 2 given as two sets of locations of individuals  $\x_1=\{x_{11},...,x_{1n_1}\}$ and $\x_2=\{x_{21},...,x_{2n_2}\}$ respectively, where the locations are observed in a well--defined area. We will call the combined set of points $(\x_1,\x_2)$ a bivariate point pattern, and refer to the individuals' locations simply as points. Technical details are left to Appendix \ref{A:details}, but in brief we assume that the data generating mechanisms can be described by some processes $X_1$ and $X_2$, and the goal of statistical analysis is to draw conclusions about the processes using the observed set $(\x_1,\x_2)$. We start by assuming that the processes are second--order stationary, which means there is no underlying heterogeneity in the abiotic environment (e.g. elevation, soil chemistry) that also affects the distributions of the species, and implies that the statistics calculated from the data do not depend on any particular location in the observation window (see the Discussion for extensions). Although ecological communities are rarely well approximated by stationary models, we motivate studying the stationary case as this must be explored first, before any more complex scenarios can be understood.

%
%


We will focus our attention on the second--order statistic commonly known as Ripley's $K$ \citep{Ripley1979} and its derivative, the pair correlation function; our rationale being these two summaries are amongst the most popular when characterizing joint dependence 
\citep{Perry2006, Law2009, Velazquez2016}. First (as is standard) we define the intensity of a point process $\lambda>0$ as the expected number of points per unit area. The cross-$K$ or partial-$K$, denoted here by $K_{12}(r)$, is a function defined as the expected number of points of species $2$ in a circle of radius $r$ placed on a random individual of species $1$, scaled with intensity $\lambda_2$ to remove dimension and facilitate comparisons. Due to symmetry, it follows that $K_{12}(r)=K_{21}(r)$. The parameter $r$ controls for spatial scale and allows for multi-scale analysis. 

The derivative of $K_{12}$ in $r$ is denoted by $g_{12}(r)$, and is called the cross- or partial- pair correlation function (pcf). The pcf describes the aggregation/segregation of cross species point locations at distance $r$ where the probability of having a species 1 individual in some small region and a species 2 individual in some small region distance $r$ away is relative to $g_{12}(r)\lambda_1\lambda_2$. The quantities are scaled so that for independent processes the expectation is $K_{12}(r)=\pi r^2$ and $g_{12}(r)=1$. The different statistics are used to ask subtly different questions, with $K_{12}(r)$ testing for species independence \emph{up-to} distance $r$, and $g_{12}(r)$ testing for independence \emph{at} distance $r$.

\subsection{Model generated data for illustration}

For better understanding of the power of bivariate point pattern statistics, we develop a simple two-species model for which the level of cross-species aggregation/segregation can be controlled directly and explicitly by two parameters that determine the spatial scale and the strength of the interaction. Using this model we can provide power estimates for different sample sizes and interaction scales and strengths using simulations. 
The details of the model are provided in Appendix \ref{A:model}. Briefly, we assume species 1 is insensitive to the presence of species 2, but that the spatial distribution of species 2 is dependent on the spatial distribution of species 1. Asymmetric interactions are a reasonable starting point given they are thought to be quite common in plant communities 
\citep{Freckleton2002} and theory suggests competitive asymmetry may help maintain diversity in competitive communities 
\citep{Nattrass2012}. The locations of all $n_1$ individuals are given by a Poisson process so species 1 exhibits no intraspecific spatial structure. The $n_2$ individuals are placed with distribution that depends on the locations of species 1. 
Importantly the model has 
\[
g_{12}(r) = 1 + b h(r),
\]
where $h(r)=\exp[-r^2/(2\rsigma^2)]$ is a decreasing function whose exponential decay is controlled by the parameter $\rsigma>0$, and has a range ($h$ is non-zero) of approximately $2\rsigma$. This function is analogous to the interaction or competition kernels used in spatially explicit birth-death models 
\citep{Murrell2003, Murrell2010}. The strength of interspecies' interaction, as summarized by $g_{12}(r)$, is controlled by the parameter $b\ge-1$. If $-1<b<0$ the two species exhibit segregation $(g_{12}<1)$, if $b>0$ the two species exhibit aggregation or clustering $(g_{12}>1)$, and when $b=0$ the two species are independent. The reader should note that this model is simply a pattern generating process for illustration, rather than a mechanistic model, and we simulate patterns conditional on fixed $n_1$ and $n_2$ as we want full control over them (for the unconditional model the abundances are random, like in the birth and death processes, see e.g. \citealt{Murrell2010}). 
Example point patterns showing inter-species aggregation and segregation can be found in Appendix \ref{A:model}, Fig. \ref{fig:A1}. 




\subsection{Testing bivariate independence}	+
We now turn our attention to the main problem of determining  if the processes $X_1$ and $X_2$, as observed through the bivariate point pattern $(\x_1,\x_2)$, are statistically independent. If the processes were independent, then the observed pattern would be a \emph{random superposition} of the two processes. We will take this as our \emph{independence}  or \emph{null hypothesis} 
which now needs to be tested using the observed data.

To test if the independence hypothesis is compatible with the data, observed values of a chosen test statistic are compared to the distribution of the test statistic under the independence model. We can test either a) at some specific range, which we call \emph{pointwise tests} or b) simultaneously over multiple ranges. For both types of tests the idea is to compute some test statistic $T\in\R$ from the data, and compare it to the values of $T$ (its distribution) as if the null hypothesis were true. If the data value is sufficiently extreme, we have reason to reject the null hypothesis.

The true distribution of the test statistic under independence is rarely known in point pattern applications, and needs to be approximated by an empirical distribution derived from simulations under the independence model. This approach is known as Monte Carlo testing \citep{Myllymaki2017}. We consider the observation area to be rectangular, in which case the independence simulation consists of randomly shifting pattern 1 (or 2 or both) with a toroidal wrap \citep{Lotwick1982}. This keeps the intra-species statistics of the patterns intact while "breaking" any inter-species dependencies, and can also be used for inhomogeneous patterns \citep{Cronie2015}. 



For the purposes of this discussion, we will consider only the simple pointwise testing scenario, for which we can employ an analytical approach using a Gaussian approximation to the distribution corresponding to the random shift simulations. As we will show, the approximation is very useful since it is not only computationally very efficient relative to the MC simulations, but also allows some analytical insight into what affects the power of the tests. The pointwise tests we will study are comparable to simultaneous tests when the best range to test at is known (see Table \ref{tab:dev2} in Appendix \ref{A:power}).  As detailed in Appendix \ref{A:details}, we can choose an unbiased estimator $\hat K_{12}$ for which approximately it holds:
\begin{align}
\label{eq:gaussian}
\hat K_{12}\sim N(K_{12}, \sigma^2)\quad \Longleftrightarrow\quad T := \frac{\hat K_{12} - K_{12} }{\sigma} \sim N(0,1),
\end{align}
where $K_{12}$ is the value under the correct model. Conditional on the observed point counts $n_1,n_2$, the variance of $\hat K_{12}(r)$ can be approximated by
\begin{equation}
\label{eq:var}
\sigma^2 \approx c_1 (n_1 n_2)^{-1}[c_2(n_1+n_2)+c_3],
\end{equation}
where $c_1,c_2,c_3$ are constants depending on the range $r$ and the geometry of the observation area (see Appendix \ref{A:details}). The variance is exact when $X_1$ and $X_2$ are uniformly distributed, but works quite well also for internally aggregated/segregated patterns as we will see later on in Section \ref{sec:balanced}. Although we focus on $K_{12}$, the approach is nearly identical for $g_{12}$, only the constants are different.



%

\subsection{Power of a statistical test}
Denote the null hypothesis of bivariate independence by $H_0$, the test statistic by $T$, and a confidence level of the test by $(1-\alpha)$ where $\alpha\in(0,1)$. Recall that $\alpha$ is the researcher's fixed accepted margin of making a false positive decision, also known as \emph{type I error}, defined mathematically as
\[
\alpha \ge P_0(T > q_{1-\alpha}|H_0 \text{ true}),
\]
where $P_0$ is the distribution of $T$ when $H_0$ is true, $q_{1-\alpha}$ is the corresponding threshold value for $T$ so that if $T>q_{1-\alpha}$ under $H_0$ we reject the null hypothesis $H_0$. The condition refers to $T$ being tested. On the other hand, the \emph{power} of a test is the probability of a true positive judgment, i.e. the probability of rejection when the hypothesis $H_0$ does not hold. Consider first the margin of making a false negative judgment,
\[
\beta \ge P_0( T \le q_{1-\alpha}|\ H_0 \text{ not true}),
\]
also known as \emph{type II error}. Then the power of the test is defined as
\[
power = power(H_0,  T, \alpha) := 1-\beta .
\]
Therefore, a test is powerful if it can correctly reject the wrong null model with a high probability. 

Consider the idealized situation of testing the cross-species independence using the pointwise summary $K_{12}=K_{12}(\tilde r)$ for some fixed spatial scale $\tilde r$ only. For the test statistic $K_{12}$ the null hypothesis $H_0$: 
``random superposition'' implies $K_{12}=k_0=\pi \tilde r^2$. Let us now consider the situation that in truth $K_{12} = k_{12} \neq k_0$. Then if we accept the approximate Gaussianity of the test statistic as shown in the previous section, it follows by elementary manipulations that
\begin{align}
\label{eq:power} 
power \approx & 
1-\Phi\left(q_{1-\alpha} - \frac{|k_{12}-k_0|}{\sigma}\right),
\end{align}
where $\Phi$ is the cumulative distribution function of the standard Normal distribution, with $a$-quantiles $q_a$. Notice that the sign of interaction does not matter, meaning that due to symmetry of the Gaussian distribution aggregation is as easy or hard to detect as segregation of similar strength. Also notice how the power is dependent on the variance ($\sigma^2$) of the test statistic used. The smaller the variance, the higher the power, which explains why different unbiased estimators of $K_{12}$ have been developed \citep[see e.g.][]{Illian2008a}
and, while all being correct in the sense of bias, they can lead to different rates of detecting interactions because of different variances. 

We can now use the power formula and our approximation for the variance (equation~\ref{eq:var}) to illustrate how to
\begin{itemize}
	\item compute the power of the test given the point counts $n_1,n_2$, expected true signal $k_{12}$, and the type I error tolerance $\alpha$;
	\item compute the required point counts given the expected true signal $k_{12}$, the type I error tolerance $\alpha>0$ and the type II error tolerance $\beta>0$ or power.
\end{itemize}

%

\section{Results}
The power formula (equation \ref{eq:power}) is a good approximation to the power of the toroidal shift Monte Carlo test (Fig. \ref{fig:fig2}). There is very little difference between the test's true power and the approximative power given by the analytical formula, with the analytical approximation slightly overestimating the power (at most~$10\%$). This implies that we can discuss the power and its effect on ecological interpretations using the convenient analytical formula, acknowledging the small optimistic bias.

As indicated by equation~\eqref{eq:var} the variance of the estimator for the $K_{12}$-function is increased when either or both of $n_1$ and $n_2$ are small. This means that both the imbalance in population abundances as well as the total number of individuals affect our ability to detect bivariate interactions. We shall investigate each of these in turn, as well as the spatial range of testing.

\subsection{Power in balanced scenarios and the importance of the spatial scale of testing}
\label{sec:balanced}

\begin{figure}[!h]
	\centering
	\includegraphics[width=\textwidth]{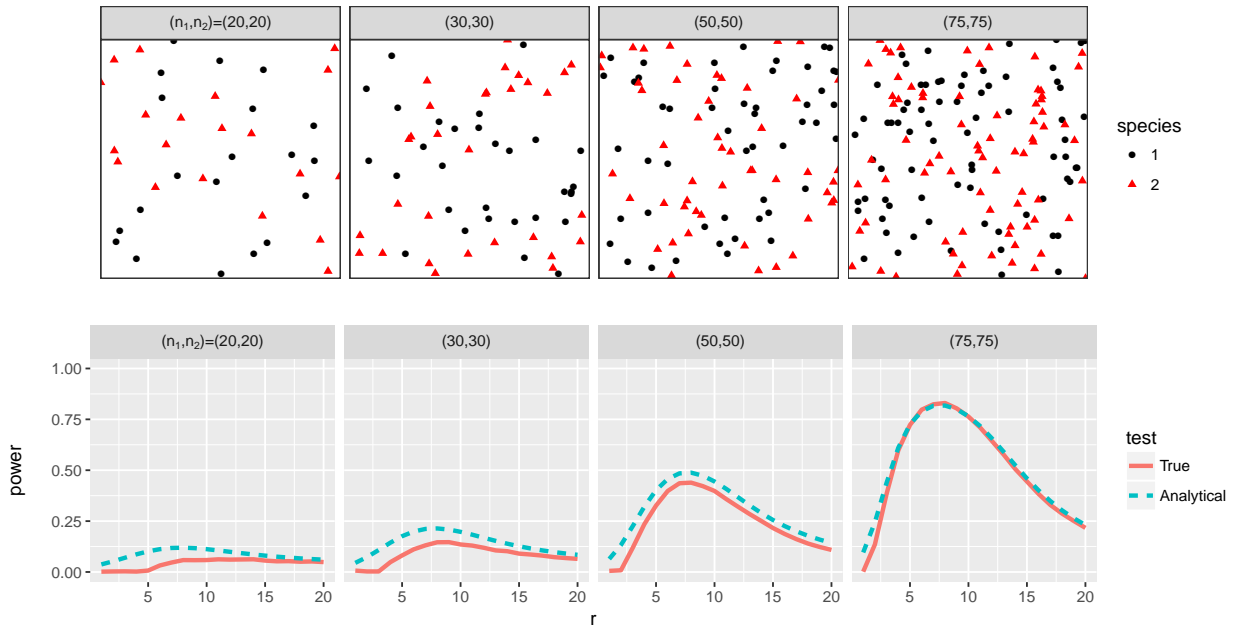}
	\caption{{\it Top:} Examples of the segregated bivariate point patterns,  $b= -0.5$ and $2\rsigma = 10$, $100\times 100$ window. {\it Bottom:} The power of $K_{12}$-based pointwise cross-species independence tests when species are segregated like in the example patterns. The true power is estimated using 5000 repeated tests with 199 random shifts each.}
	\label{fig:fig2}
\end{figure}

Fig. \ref{fig:fig2} depicts the pointwise powers for different balanced ($n_1=n_2$) low-abundance scenarios when data is segregated (aggregated results are nearly identical). Visual inspection of the example point patterns (Fig. \ref{fig:fig2}, top row) already gives some indication that departures from spatial independence might be hard to detect for the lowest abundances. More formal analysis of the power quantifies the increase in ability to detect interactions with increasing abundances ($n_1$, $n_2$) of the species being investigated and how this is affected by the spatial scale at which the hypothesis is tested (Fig. \ref{fig:fig2}, bottom row). In all cases the power to detect the interaction at small spatial scales ($r<2$) is low because, although the interaction is at its strongest here, the variance of $K_{12}$ is relatively high and overwhelms the ecological signal. The trade-off between signal and noise leads to a unimodal relationship between power and the neighbourhood radius $r$, with the peak being approximately at $r = 7$ for the interaction range $2\rsigma = 10$ for all abundance sizes considered (Fig. \ref{fig:fig2}). We will refer to this peak in power with $r$ as the optimal range for testing, 
and will focus on this best case scenario for the results presented below. The unimodal relationship highlights the point that having some prior knowledge about the likely ranges of biotic interactions is going to be important for detecting interactions.


\begin{figure}[!h]
	\centering
	\includegraphics[width=\textwidth]{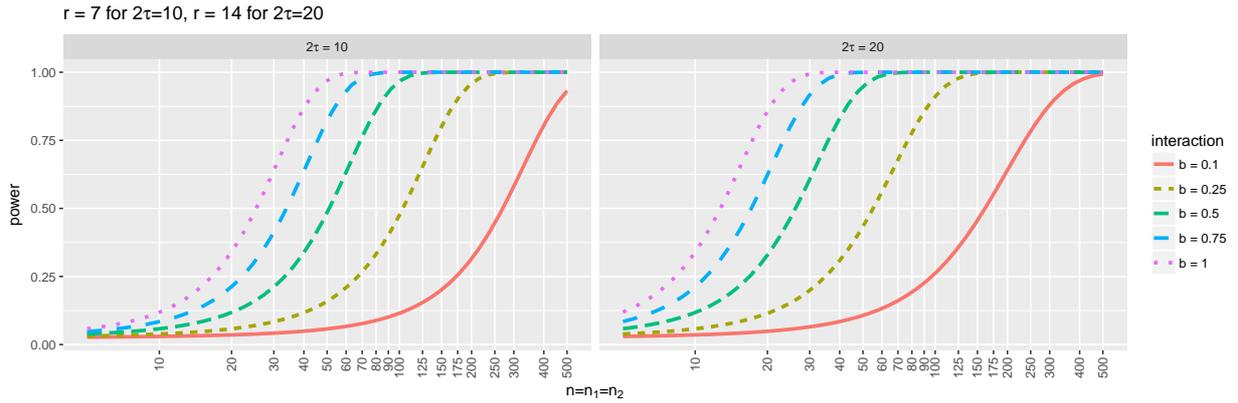}
	\caption{Power of $K_{12}$-based pointwise cross-species independence tests when abundances are balanced and testing is done on the best possible range.  Test level $\alpha=5\%$.}
	\label{fig:fig2b}
\end{figure}

Previous results based on \textit{in situ} data analysis suggest detectable interactions between trees typically occur over 10-20m~\citep{uriarte2004spatially}. Scaling our analyses accordingly, we can use the power formula to estimate the population sizes we require in order to reliably detect an interaction of a given strength and range (Fig. \ref{fig:fig2b}). If for example we wish to be 75\% sure a true positive is not to be missed when the interaction strength is weak ($b$= -0.1), then we require species with populations of approximately 400 individuals for the 10 unit  interaction neighbourhood ($2\rsigma$ = 10) and 250 individuals for 20 unit neghbourhood ($2\rsigma$ = 20). This value is surprisingly large compared to what data we commonly have available to us.

In contrast, for the maximum possible negative interaction strength ($b$= -1), a similar level of power is reached with only around 35 individuals for $2\rsigma$ = 10 unit and 18 individuals for $2\rsigma$ = 20. Conversely, if we have a pair of species with $n_1$=$n_2$ = 50, and we wish to be 75\% sure a true positive is not missed, we must hope that the true interaction $|b|$ when coupled with short interaction range ($2\rsigma=10$) is at least $0.7-0.75$, and if coupled with long interaction range ($2\rsigma=20$) is at least $0.3-0.4$. It therefore seems likely that only the very strongest interactions are detectable with the number of individuals that are typically found in the species-rich datasets.
 

\subsection{Imbalances in species abundance}
\begin{figure}[!ht]
	\centering
	\includegraphics[width=\textwidth]{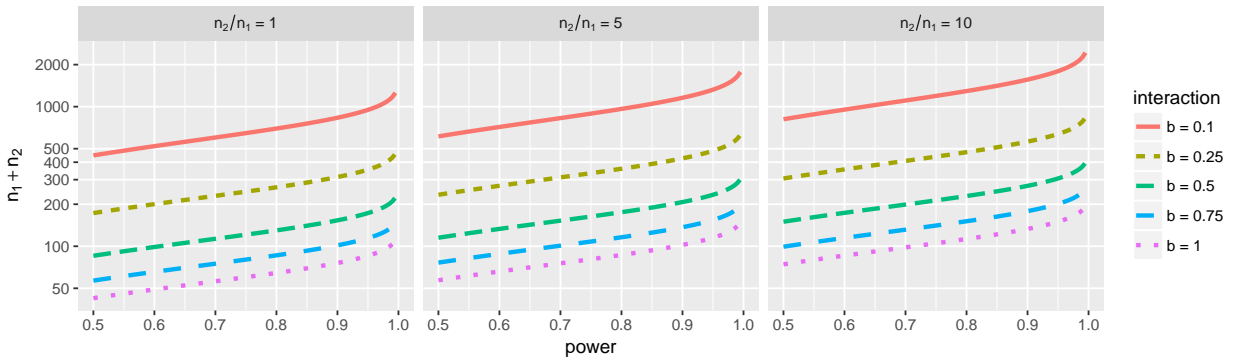}
	\caption{Sample size $n_1+n_2$ requirements if testing for independence at level $\alpha = 5\%$ with a $K_{12}$-based pointwise cross-species independence test in the example scenario. Interaction range $2\rsigma=10$.}
	\label{fig:fig5}
\end{figure}

Since most communities exhibit a `hollow curve' distribution of population abundances  \citep{McGill2007}, 
an imbalance in population sizes is very common. From the variance formula \eqref{eq:var} it is clear that imbalance has a strong effect on the power because the term $(n_1 n_2)^{-1}$, and hence the variance, increases with imbalance. 
This relationship is confirmed when we use the power formula to quantify the effect of population imbalance for different interaction strengths and combined population sizes (Fig. \ref{fig:fig5}). So, for example, for an interaction strength of $|b|$ = 0.1 and a desired power of 80\%, a combined individual count of about 750 is required when the populations are perfectly balanced, but 1000 are required when one species is five times more abundant than the second species, and a surprisingly large 1500 required when one species is ten times more abundant than the other. Alternatively, consider that we require 90\% power, and that the interactions are assumed to be $|b|=0.5$ and of short range, $2\rsigma=10$. Then, to be on the safe side, we should attain samples of sizes at least $(100,100), (40,200)$ or $(30,300)$, depending on the imbalance. 




\subsection{Power at rainforest sample sizes}
We now consider how our understanding of the power to detect interactions might affect results for observed plant communities. For simplicity, let us assume interactions are of the type given by our model 
and that every species is interacting with every other species in an identical manner (so $b$ and range $2\rsigma$ are the same for all pairs of species). Since the power is the probability of detecting interactions, we can get a rough estimate of the number of detected cross-species interactions by assuming the tests are independent, and summing up the powers. This then allows a coarse comparison of recently reported frequencies of detected interactions in tropical forests \citep{Wiegand2012b,Wang2014,Perry2014,Lan2016,Chacon-Labella2017} with the expected frequency of detected interactions as a function of power.



\begin{figure}[!ht]
	\centering
	\includegraphics[width=1\linewidth]{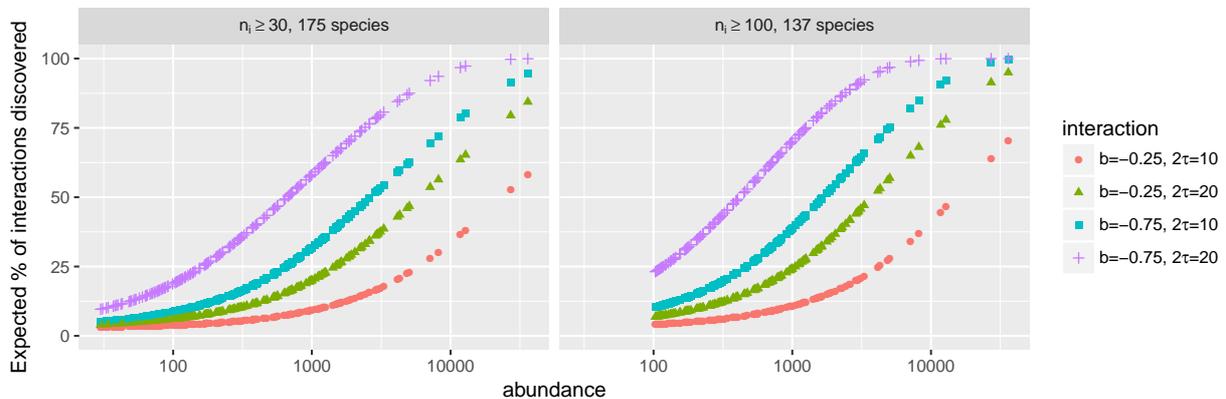}
	\caption{Expected fraction of interaction detected per abundance, if all pairs of species were to interact, and we tested independence with $K_{12}$ at optimal range.}
	\label{fig:fig6}
\end{figure}

Fig. \ref{fig:fig6} shows the expected number of cross-species interactions detected as a function of abundance for various hypothetical interaction strengths and ranges. The abundances are taken from the Barro Colorado Island (BCI) 1995 census\footnote[1]{http://ctfs.si.edu/webatlas/datasets/bci/abundance} of woody plants with diameter at breast height at least 1cm \citep{Condit1998}. 
The abundances are highly skewed, with a large proportion of low abundance species, and we show the power in two cases, when the pool of species consists of those with abundance at least 30 and 100. Reducing the species pool by increasing the abundance threshold naturally increases the proportions of detection, and highlights the importance of using similar thresholds when comparing different communities. It is striking how little power is to be expected for most of the species even when assuming strong interaction ($b$ = -0.75). Only when the abundance of a species reaches thousands, can we be expected to detect even 50\% of the interactions present. This is a very thought-provoking result, as the lack of detection might be explained simply by a lack of power in the majority of species-pairs.

Making the same assumptions about identical interactions between all pairs of species, we conducted a comparative analysis using abundances of plants 
in the BCI 1995 census, the Changbaishan (CBS) forest plot \footnote[2]{https://doi.org/10.5061/dryad.p4n8rg64} ~\citep{wang2010spatial}, 
and the Sinharaja 1995 census \footnote[3]{http://ctfs.si.edu/site/Sinharaja/abundance} \citep{Wiegand2007a}. Following \cite{Wiegand2012b}, we considered only large plants (diameter $\ge$10cm) and keep only those species with at least 70 large individuals (note that \citealt{Wiegand2012b} thresholded CBS at 50 individuals). The analysis shows a big difference in the expected proportion of interactions that would be detected due to differences in the species abundances of the communities (Table \ref{tab:forests}). Although the actual proportions differ, qualitatively, these results are in a sense similar to those reported by \cite{Wiegand2012b} who detected that approximately $70\%, 60\%$ and $30\%$ of interactions departed from their null hypothesis in the CBS, Sinharaja and BCI communities, respectively. 

\begin{table}[ht]
	\centering
    \begin{tabular}{r|ccc}
	& \multicolumn{3}{c}{Expected proportion of significant interactions}\  \\
	Interaction model & CBS (13/52) & Sinharaja (47/207) & BCI (61/305) \\ 
		\hline
  $b=-0.25, 2\tau=10$ & 0.46 & 0.19 & 0.12 \\ 
  $b=-0.25, 2\tau=20$ & 0.71 & 0.42 & 0.29 \\ 
  $b=-0.75, 2\tau=10$ & 0.84 & 0.62 & 0.46 \\ 
  $b=-0.75, 2\tau=20$ & 0.96 & 0.91 & 0.80 \\ 
 \hline
	\end{tabular}
\caption{Expected proportion of significant $K_{12}$ independence tests at most powerful range and type I error level $\alpha=5\%$, in three different forest communities. In parenthesis: species considered in the analysis / total species richness. Analysis includes only species with at least 70 large (diameter$\ge$10cm) individuals, and abundances of only those large individuals.}
\label{tab:forests}
\end{table}

These toy examples highlight how the relationships of the power to detect interactions with population sizes, strength of interaction and spatial scale of interaction can in principle lead to patterns similar to those described in previous studies. Fortunately these artificial examples can be taken as ``worst case'' scenarios. Many of the species in forest plots are highly localised to environmental niches 
\citep{Harms2001,flugge2014method}, 
in which case the context of testing needs to be defined more accurately and the pool of potential interactions limited, thus regaining power.


  
\section{Discussion}



Understanding the relative strength and therefore the importance of interspecific interactions is one of the key goals of community ecology, and the null model approach has been popular for characterizing spatial point patterns of (predominantly) diverse plant communities \citep[e.g.][]{Martinez2010,Wiegand2012b,Wang2014,Perry2014,Lan2016,Velazquez2015, Chacon-Labella2017,Wiegand2017}. However, there has been little guidance on when a given test is likely able to detect species associations that are present. Here we have made a first step in closing this important gap in our understanding. Our results clarify the quantitative relationships between the strength of the underlying biological interaction, sample size (number of individuals of both species under investigation), and the spatial scale over which the test is being performed. We have also shown that statistical power may explain both the low detection rate of biological interactions in plant communities, and the negative relationship between species-richness and frequency of detected interspecific interactions in comparative studies. 

Ecologists have had to rely largely upon their intuition for deciding the minimum population size to include in their analyses with the result that a range of criteria up to 100 individuals~\citep{flugge2014method} have been used. For species-rich communities, where many interspecific interactions may necessarily be weak \citep{Chesson2000}, abundances of both species may need to be in the hundreds of individuals before any interaction is detected (Fig. \ref{fig:fig5}), and this implies previous abundance thresholds are likely too low to detect many interactions. As several authors have acknowledged, the failure to reject the null hypothesis of spatial independence in so many species-pairs does not necessarily mean interspecific interactions are not occurring, or present \citep{Wiegand2012b,Perry2014,Chacon-Labella2017}. We hope our study highlights how the power of the tests can be assessed and should be factored into the interpretation of the results. The power formula can also be used in estimating the area of observation necessary to increase the power to a desirable level (Appendix \ref{A:Area}), so can also be used to aid study design. Despite this, we do stress that there is still much to be learned about the power of the statistical tests used in earlier studies, given the assumptions we had to make, and that the reader should take our contribution as a first step that offers a rough guide to sample sizes that are required to make strong statements about the frequency and strength of interspecific interactions. 



Although our model is clearly mis-specified as we use tests assuming that intensity is not dependent on abiotic features of the environment, the general applicability of our results will carry-over into the inhomogeneous setting. In particular we would still expect a positive relationship between population size and frequency of interactions to emerge simply due to an increase in power at larger sample sizes. Such a positive relationship has already been reported in a number of empirical studies that take habitat associations into consideration \citep{Wiegand2012b,Luo2012,Wang2014,Chacon-Labella2017}. It is possible that common species are better competitors and are somehow suppressing the abundance of the weaker competitors, but without experimental manipulation, or perhaps different analyses using repeated sampling over time~\citep{damgaard2017s}, it is hard to distinguish whether this pattern is a result of biological processes or the ability of the statistical methods to detect interactions at different population sizes.

The spatial scale over which tests are performed is important for the ability to detect spatial dependencies (Fig. \ref{fig:fig2}), and our results are similar to empirical studies that often find few negative interactions at the shortest distances, even though this is where the interactions are likely to be strongest \citep{Wiegand2012b,Wang2014, Chacon-Labella2017}. Short scales suffer from having high variability due to the relatively small number of neighbours possible in a small area, but at longer distances, the effect of neighbours is weaker. Hence there is a sweet spot where this trade-off is maximised, and the location of this is likely dependent on several factors, not least of which is the scale over which interactions are occurring (e.g. Fig. 2 in \cite{Chacon-Labella2017} for an empirical example). For woody plants, there have been several studies that have fitted neighbourhood growth or survival models to individual-based data that tracks the fate of trees over time \citep[e.g.][]{uriarte2004spatially, Stoll2005}, and most results seem to point to interactions being confined to 10-30m radius around an individual. However, little is known about how the spatial scales of interspecific interactions change with life history stage, environmental conditions, or even species identity even though the latter has been shown to be very important for determining coexistence ~\citep{Murrell2003}. Any changes to the scales of interactions will have consequences for the hypothesis testing methods discussed here, but until more is understood about the spatial scales of interactions between species, it seems sensible to test over ranges reported in earlier studies.

Our discussion up to this point has been in the context of stationary, most notably homogeneous, data. Most recent analyses have tried to factor out the effects of spatial heterogeneity in the abiotic environment by using inhomogeneous Poisson processes as the null model \citep{Wiegand2012b,Punchi-Manage2015, Chacon-Labella2017}. Currently it is hard to predict whether the power of an inhomogeneous analogue of our scenario would be lower or higher. On the one hand we could expect higher power to detect interactions because the model better captures the underlying processes that generate the spatial distributions of the species within the community. However, we also expect variance to be increased, since extra parameters need to be estimated leaving a fewer degrees of freedom per parameter. For example, tests using the inhomogoneneous Poisson process method use a smoothing kernel to approximately remove the effects of large scale structure assumed to be caused by habitat associations (see e.g. \citealt{Wiegand2012b}). Typically, the same smoothing parameter is used for all species, which is a sensible assumption when little is known about the spatial scale of habitat associations, but there is no reason to suspect a single smoothing parameter is appropriate for all species. 
An open challenge is to better understand how mis-specification of the smoothing parameter will bias the detection of interactions. Again, we feel that using a biologically motivated model to simulate data is a useful approach for exploring such issues.


Finally, we remind the reader that the spatial statistics used in the null model approach do not say anything directly about the processes that may have created the patterns, and different processes could generate the same summary statistic. As an alternative, model-based approaches, either in the form we use here (which include the familiar Thomas Cluster models) or birth-death models \citep{May2015, rajala2017detecting} could also be applied to the inference of biological interactions from point pattern data \citep{Wiegand2017}. Model fitting will normally lead to estimation of parameters that can also be estimated in the field (eg. dispersal kernels, interaction kernels), we therefore feel that their continued development will help improve understanding of the processes underpinning the results returned \citep{Wiegand2017}.

In conclusion, we hope our main contribution is to encourage more users to consider explicitly the ability of the spatial point pattern tests to detect significant associations between species. 
We have shown that the data requirements to detect even strong interactions may be quite high, mirroring results for null model tests of species co-occurrences in community matrix data 
\citep{Gotelli2000,Freilich2018}. On this basis, we suggest it is desirable to only interpret the frequency of interactions across large numbers of species once the effect of different powers to detect interactions for pairs of species of given population sizes has been (even approximately) factored out. This seems especially important in comparative analyses across different communities where the spatial scales, strengths of interactions and the species abundance distributions may differ and affect the power to detect interactions.

\section{Authors' contributions}
TR conceived the idea during discussions with DM and SO; TR derived the model and formulas, designed and executed computations, and contributed extensively to the manuscript; DM contributed to the rainforest experiment and extensively to the manuscript; SO contributed to the manuscript. All authors contributed to the intellectual core of the manuscript. 


\section{Acknowledgements}

The work has been supported by UK Engineering and Physical Sciences Research Council (EP/N007336/1) and European Research Council (CoG 2015-682172NETS).

\bibliographystyle{plainnat}
\bibliography{biv-power}

\newpage

\appendix
\renewcommand{\thefigure}{S\arabic{figure}} 
\renewcommand{\thetable}{S\arabic{table}}

\section{Technical details on point processes and summaries}
\label{A:details}
\subsection{Preliminaries}

We observe two point processes $X_1$ and $X_2$, observed as point patterns, or equivalently as sets of locations, $\x_1=\{x_{11},...,x_{1n_1}\}$ and $\x_2=\{x_{21},...,x_{2n_2}\}$ in a finite observation window $W\subset \R^2$. Write $N_i(B):=\#(X_i\cap B)$ for the random number of points of type $i$ in a set $B\subset \R^2$. All results generalise easily to higher dimensions. We assume that the processes $(X_1,X_2)$ are jointly second order stationary, so that the expectation of the statistics we shall calculate do not depend on any particular location in the observation window.


\subsection{Summary statistics for bivariate interaction}

First assumption is that the expected point count in any set $B$ can be written as an integral
\[
\E N_i(B) = \int_B \lambda_i(u)du,
\]
where $\lambda_i(u)\ge 0$ is called the intensity. For stationary processes $\lambda_i(u)\equiv \lambda_i$ is a constant, and we assume that $\lambda_i>0$, so that $\E N_i(B)=\lambda_i |B|$.

We define the cross-$K$ function as a function of a range parameter $r>0$  
\[
K_{12}(r) := \lambda_2^{-1} \E_{o1} N_2(b(o,r)),
\]
where $b(o,r)$ is a ball of radius $r>0$ centred at the origin, the expectation $\E_{o1}$ is conditional on the joint process having a point of type 1 at the origin $o$ (for stationary processes the exact location does not matter). Heuristically, $\lambda_2 K_{12}(r)$ is the mean abundance of species 2 within distance $r$ of a typical point of species 1. Equivalently we can define $K_{21}(r)$, but due to symmetry $K_{12}=K_{21}$.

The cross-$K$ index is a powerful statistic for testing purposes, but for a more detailed description of spatial interactions we often study the derivative of $K_{12}$, 
\[
g_{12}(r):=\frac{K_{12}'(r)}{2\pi r},
\]
known as the cross (or partial) pair correlation function (pcf). The pcf describes the aggregation/segregation of cross species point locations: The probability  of having a species 1 point at some small region $dx$ and a species 2 point at some small region $dy$ is given by $g_{12}(||x-y||)\lambda_1\lambda_2 dx dy$. If the processes are independent, $g_{12}(r)\equiv 1$ and $K_{12}(r)=\pi r^2$. We say the processes are aggregated if $g_{12}> 1$, and segregated if $g_{12} < 1$, at any particular range $r>0$.  

To estimate these quantities several estimators have been proposed, differing in how the observation bias near the borders of $W$ is corrected. We will look at bivariate versions of the globally corrected "Ohser"-type estimators  (\citealtApp{Illian2008a}, p. 230, \citeApp{Ward1999} and \citealtApp{Wiegand2016}) of the form
\[
c(r) \sum_{x\in \x_1}\sum_{y\in \x_2} f_r(x-y) = c(r) T(r)
\]
where $c(r)$ is some constant possibly depending on $r$, and $f_r$ is some function, for example an indicator function for $K_{12}$ and a kernel function for $g_{12}$. Note that while the theoretical $K_{12}$ and $g_{12}$ are symmetric in the species indices, their estimators are not necessarily so (see e.g. \citealtApp{Lotwick1982}).

\subsection{Covariance of the summary statistics}
\label{A:details_cov}
Consider two point processes $X_1$ and $X_2$ with fixed counts $n_1,n_2$ in a window $W$. We are interested in the covariance between different ranges $r>0,s>0$ of estimators of type 
\[
M(r) = c(r) \sum_{x\in X_1}\sum_{y\in X_2} f_r(x-y) = c(r) T(r) ,
\]
where $f_r$ is symmetric in $x-y$ (this can be easily extended to non-symmetric functions, see \citealtApp{Lotwick1982}). If $X_1$ and $X_2$ are sets of independently and uniformly distributed points in $W$, then 

\begin{align*}
Cov[T(r),T(s)] &=(n_1n_2)^{-1}[ (n_1+n_2-2)c_1(r,s) + c_2(r,s) - (n_1+n_2-1)c_3(r,s)]\\ 
&= (n_1n_2)^{-1}\left\{(n_1+n_2)[c_1(r,s)-c_3(r,s)] + c_2(r,s) + c_3(r,s) - 2c_1(r,s)\right\},
\end{align*}
where

\begin{align*}
c_1(r,s) &= \E f_r(x-y)f_s(x-z) \\
c_2(r,s) &= \E f_r(x-y)f_s(x-y) \\
c_3(r,s) &=\E f_r(x-y)\E f_s(x'-y'),
\end{align*}
with expectations over i.i.d. uniform random variables $x,x',y,y',z$ on $W$. 

For the box-kernel estimator of the pair correlation function $g_{12}$, we choose $$f_r(x-y)=(2h)^{-1}1_{b_r}(x-y),$$ where $b_r=b(o,r+h)\setminus b(o,r-h)$ and $h>0$ is the bandwidth. Then

\begin{align*}
c_1(r,s) &= I_3(r,s)|W|^{-3}(2h)^{-1}\\
c_2(r,s) &= I_2(r,s,h)|W|^{-2}(2h)^{-2}\\
c_3(r,s) &= [I_1(r+h) - I_1(r-h)][I_1(s+h) - I_1(s-h)]|W|^{-4}(2h)^{-2},
\end{align*}
where

\begin{align*}
I_1(r) &= \int_{b(o,r)}|W\cap W_z|dz\\
I_2(r,s,h)&=I_1(r+h)-I_1(s-h) \text{ if }|r-s|<2h, \quad\text{ and 0 otherwise}\\
I_3(r,s) &=\int_W\int_W\int_W 1_{b_r}(x-y)1_{b_s}(x-z)dxdydz.
\end{align*}

For the $K_{12}$ function $f_r(x-y)=1_{B_r}(x-y)$ where $B_r=b(o,r)$, and

\begin{align*}
c_1(r,s) &= I_4(r,s)|W|^{-3}\\
c_2(r,s) &= I_1(\min(r,s))|W|^{-2}\\
c_3(r,s) &= I_1(r)I_1(s)|W|^{-4}
\end{align*}
where

\begin{align*}
I_4 &= \int_W\int_W\int_W 1_{B_r}(x-y)1_{B_s}(x-z)dxdydz .
\end{align*}

The quantities $I_3$ and $I_4$ can be approximated numerically, and $I_1$ can be derived using the isotropised set covariance of $W$ \citep[][p. 485]{Illian2008a}.

\subsection{Gaussian approximation of $\hat{K}_{12}$}
\begin{figure}[!h]
	\centering
	\includegraphics[width=\textwidth]{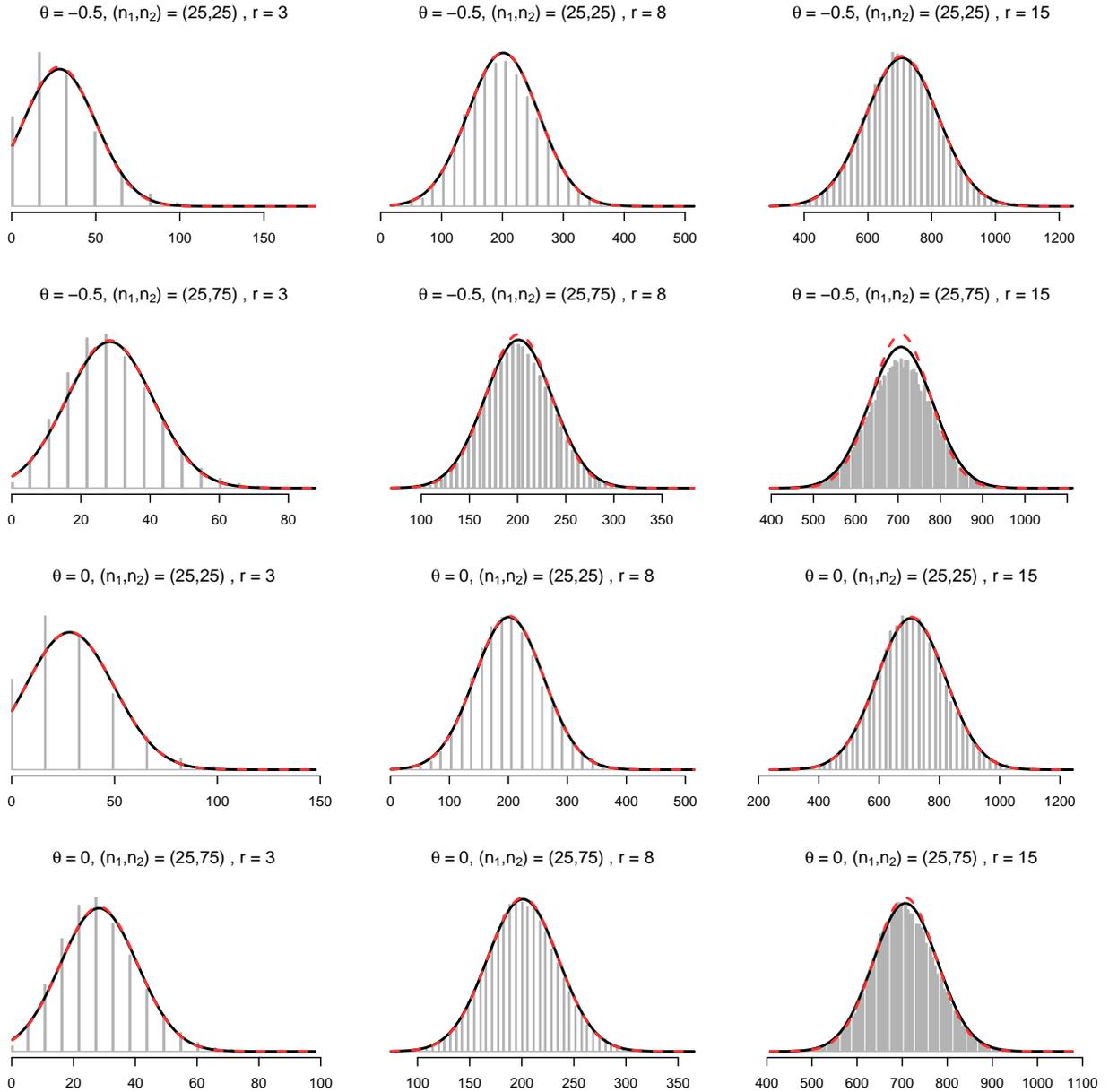}
	\caption{Empirical distribution of $\hat K_{12}$ at $r=3,8,15$ for random shifted data, where data is generated by the introduced bivariate model with range $2\rsigma=10$. Overlaid are Gaussian density functions with empirical mean and variance (solid, black) and with theoretical mean $\pi r^2$ and variance $\sigma^2$ given by the analytical approximation (dashed, red), which in most subplots is superimposed with the solid black line. }
	\label{fig:figA1}
\end{figure}

The mathematics of the limiting behaviour of the "Ohser"-type estimators are beyond this study, and for progress in this regard we refer to \citeApp{Heinrich2015}. We resort to the same argument as \citeApp{Wiegand2016}: the empirical plots do not show signs against normality apart from very short ranges due to the positivity constraint. Fig. \ref{fig:figA1} illustrates this (compare to \citealtApp{Wiegand2016}, Fig. S3). Note the accuracy of the analytical formula derived in Appendix \ref{A:details_cov} for the variance.

\section{Model generated data}
\label{A:model}

\begin{figure}
	\includegraphics[width=\linewidth]{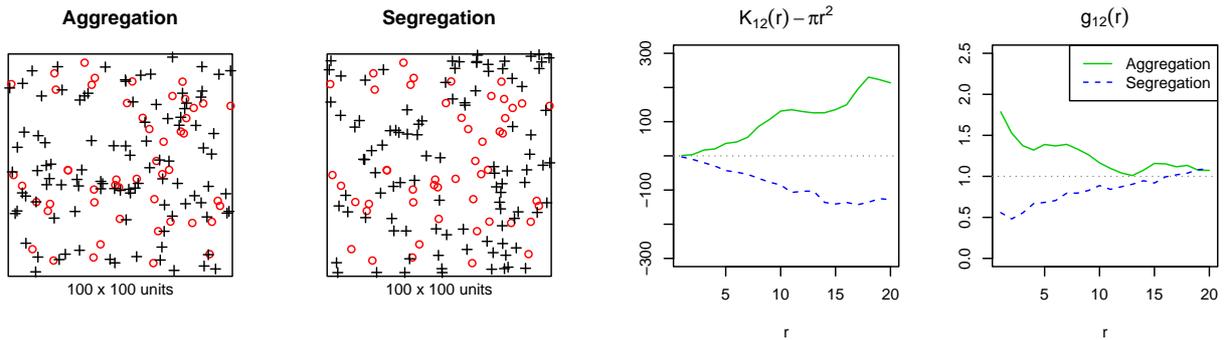}
	\caption{Example bivariate point patterns exhibiting cross-species spatial aggregation and segregation, and the corresponding cross-$K$ and cross-pcf statistics.}
	\label{fig:A1}
\end{figure}

The process is inspired by the shot-noise product Cox processes~\citepApp{Jalilian2015}, and is constructed hierarchically. First, let $X_1$ be a stationary Poisson process with intensity $\lambda_1$. Then conditional on a realisation $\x_1$ of $X_1$, let $X_2$ be an inhomogeneous Poisson process with intensity function
\[
\lambda_2(u;\x_1) = e^a\prod_{x\in \x_1}\left(1+b h(u-x)\right)\in \quad u\in W
\]
where $h(v)=k_\rsigma(v)/k_\rsigma(0)$ with $k_\rsigma$ a 2D kernel function (probability density) with standard deviation $\rsigma>0$, and $a\in \R, b>-1$ are parameters controlling the intensity and the interaction, respectively. The joint model is stationary, and isotropic if $k$ is isotropic, and has
\[
\lambda_2 = \exp\left(a+\lambda_1 b/k_\rsigma(0)\right),\ \ g_{11}(u) \equiv 1,\quad g_{12}(u) = 1 + b h(u),\quad g_{22}(u) = \exp\left(\lambda_1b^2(h*h)(u) \right)
\]
where $*$ denotes convolution. From these properties we see that if $-1<b<0$ the two species exhibit segregation $(g_{12}<1)$, and if $b>0$ the two species exhibit aggregation or clustering $(g_{12}>1)$, and when $b=0$ the two species are independent. We also see that the both types of interactions result in clustering of species 2 ($g_{22}\ge 1$).
The range of interaction (if defined via the pair correlation) is controlled by the parameter $\rsigma$. In our examples we use a Gaussian kernel, for which the range, i.e. $h$ is non-zero, is approximately $2\rsigma$. Fig. \ref{fig:A1} shows two examples of the process with identical type 1 patterns, together with their $K_{12}$ and $g_{12}$ estimates.

\section{Additional power estimates}
\label{A:power}
\begin{figure}[!h]
	\centering
	\includegraphics[width=\textwidth]{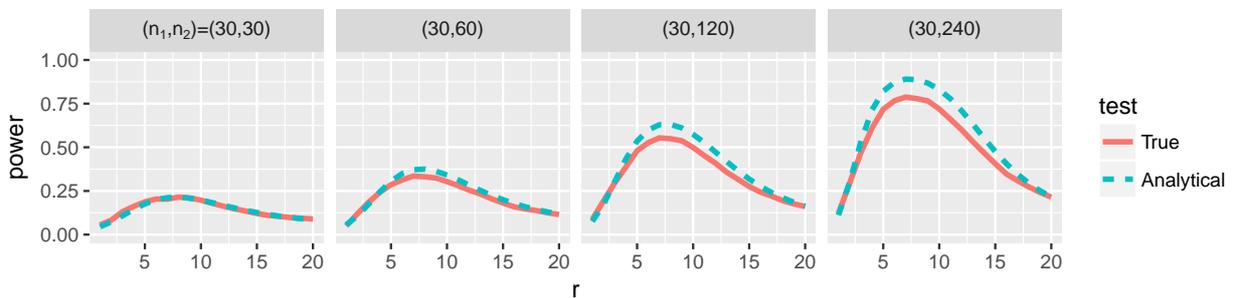}
	\caption{Power of $K_{12}$-based pointwise cross-species independence tests with varying degrees of imbalance $n_1=30\le n_2$. Range of interaction $2\rsigma=10$. The true power is estimated using 5000 repeated tests with 199 random shifts each.}
	\label{fig:fig3}
\end{figure}

Fig. \ref{fig:fig3} provides evidence that the analytical power formula is close to the true power, which can be estimated by Monte Carlo simulation, also in unbalanced scenarios. Compare Fig. \ref{fig:fig3} to Fig. \ref{fig:fig2}.

\subsection{Testing without simulation}

Note that the approximate Gaussianity and the approximate variance formula lead directly to a $\chi^2$-test of independence without random shift simulations, much like in the work by \citeApp{Wiegand2016}. 
Procedure:
\begin{enumerate}
	\itemsep0pt
	\item Estimate $\hat K_{12}(r)$ for one $r$
	\item Compute $\sigma(r)$
	\item Compute $T=(\hat K_{12}(r)-\pi r^2)^2/\sigma^2(r)$
	\item Compare $T$ to the $\chi^2$-distribution with 1 degrees of freedom.
\end{enumerate}

\subsection{Pointwise test vs testing over a range}

In the simulation experiments we control all factors, so we can choose the range of the pointwise test to be optimal, i.e. the range which we know the power is highest. Table \ref{tab:dev2} compares this optimal pointwise power to the power of a test where instead of a single range an interval of ranges is tested simultaneously using a deviation test \citep[see e.g.][]{Myllymaki2017}.

\begin{table}[ht]
	\centering
	\footnotesize
	\begin{tabular}{|r|rrr|rrr|rrr|}
		\hline
		& \multicolumn{3}{c|}{($n_1,n_2$)=(25,25)} & \multicolumn{3}{c|}{($n_1,n_2$)=(50,50)} & \multicolumn{3}{c|}{($n_1,n_2$)=(75,75)}\\
		Interaction model & r=1-10 &  r=1-20 & pw.o. &  r=1-10 &  r=1-20 & pw.o. &  r=1-10 &  r=1-20 & pw.o. \\ 
		\hline
		$b=-.25, 2\rsigma=10$ & 0.11 & 0.05 & 0.07 & 0.29 & 0.14 & 0.16 & 0.50 & 0.28 & 0.30 \\ 
		$b=-.25, 2\rsigma=20$ & 0.18 & 0.10 & 0.15 & 0.41 & 0.43 & 0.43 & 0.70 & 0.78 & 0.73 \\ 
		$b=-.75, 2\rsigma=10$ & 0.43 & 0.21 & 0.31 & 0.96 & 0.82 & 0.82 & 1.00 & 1.00 & 0.99 \\ 
		$b=-.75, 2\rsigma=20$ & 0.71 & 0.77 & 0.80 & 1.00 & 1.00 & 1.00 & 1.00 & 1.00 & 1.00 \\ 
		\hline
	\end{tabular}
	
	\caption{Power comparison of the Studentised $L^2$ deviation test over two range intervals (``r=1-10'' and ``r=1-20'') with the pointwise power formula at the known optimal range (``pw.o.''). The $K_{12}$ statistic, and the deviation test powers estimated using 1000 simulations per model and/or setting as indicated, with 199 random shifts each.}
	\label{tab:dev2}
\end{table}

We tried using the covariance formula to combine several ranges to a $\chi^2$-test, but the very short range asymmetry and the non-central $\chi^2$ did not immediately lead to a useful power approximation of the Studentised $L^2$ test.

\subsection{Improving power by combining summaries}
\label{appendix:combine}
\begin{figure}[!h]
	\centering
	\includegraphics[width=\textwidth]{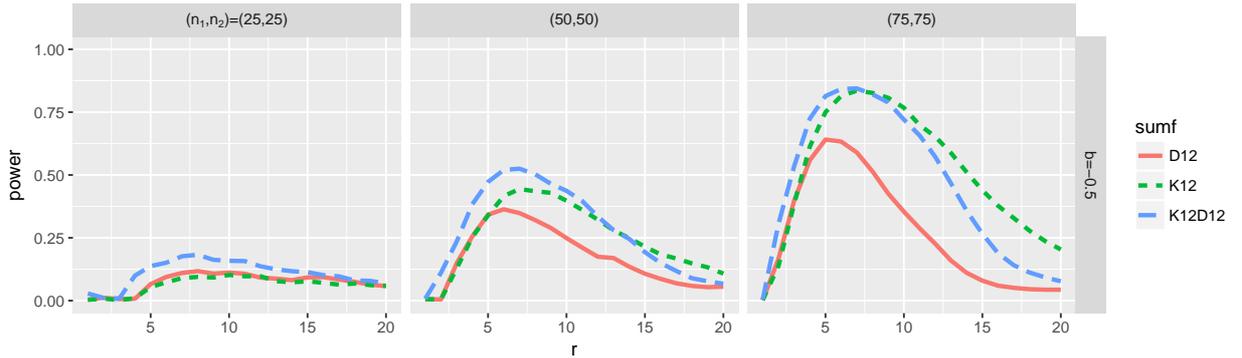}
	\caption{Power of $K_{12}$, $D_{12}$ and $K_{12}+D_{12}$-based pointwise cross-species independence tests when species are significantly segregated.}
	\label{fig:figA2}
\end{figure}

A simple way to improve power is to combine several summaries in the test statistic. As an example, we combined the $K_{12}$ with the nearest neighbour distance distribution function $D_{12}$ \citepApp{VanLieshout1999} by using the pointwise test statistic

\[
T_{KD}(r) = \left(\frac{\hat K_{12}(r)-\bar K_{12}(r)}{\hat\sigma_K(r)}\right)^2+\left(\frac{\hat D_{12}(r)-\bar D_{12}(r)}{\hat\sigma_D(r)}\right)^2.
\]
Fig. \ref{fig:figA2} depicts the pointwise powers for $K_{12}$, $D_{12}$ and the combination when the data was generated by our bivariate model with $b=0.5$, $2\rsigma=10$. The nearest neighbour summary operates only at short ranges as it saturates to 1 quickly, and for ranges $>5$ is inferior to $K_{12}$ in this scenario. But as it captures different information than the $K_{12}$, combining it with $K_{12}$ for ranges increases the power, at least when $<10$. After $r>10$ the combined pointwise power is diminished as the nearest neighbour summary provides no help yet is weighted equally with $K_{12}$ in making the decision. Weighting the statistics by their useful ranges is therefore recommended.

Table \ref{tab:A1} gives the powers of a test with $ T = \sum_{r=1}^{20} T_{KD}^2(r)$, over the ranges $1-20$. The power with the combined statistic is higher than with either of the components alone for small samples.

\begin{table}[ht]
	\centering
	\begin{tabular}{|r|c|c|c|}
		\hline
		\multicolumn{2}{|r|}{($n_1,n_2$) = (25,25)} &  (50,50) & (75,75) \\ 
		\hline
		$D_{12}$ & 0.09 & 0.27 & 0.45 \\ 
		$K_{12}$ & 0.08 & 0.43 & 0.83 \\ 
		$K_{12}+D_{12}$ & 0.14 & 0.49 & 0.82 \\ 
		
		\hline
	\end{tabular}
	\caption{Power of the independence test when using $K_{12}$, $D_{12}$ or both, for different balanced sample size, deviation test over ranges 1-20. Each power was estimated using 2000 simulations of data, 199 random shifts each.}
	\label{tab:A1}
\end{table}

\subsection{Sample size and observation window} 
\label{A:Area}



Under most circumstances samples sizes can only really be increased by increasing the area of observation, and the connection $\lambda_i \approx n_i/Area$ can be used to get a rough idea of the requirements. First, a pilot study needs to be conducted to estimate $\lambda_i$ (see \citealtApp{Illian2008a} for estimation techniques). Then we need to determine the minimum $Area$, accounting for imbalance between species if that is needed. Table \ref{tab:area} gives some example calculations when a square area is used (note that the window geometry might affect the power; see  \ref{sec:other}).

\begin{table}[!ht]
	\centering
	\begin{tabular}{|r|c|c|c|c|c|}
		\hline 
		Balance $\lambda_2/\lambda_1$ &	1 & 2 & 5 & 10 & 50\\ 
		\hline 
		Area requirement, $\lambda_1\approx 0.01$ &	$130^2$ & $110^2$ & $90^2$ & $77^2$ & $55^2$\\
		\hline 
		Area requirement, $\lambda_1\approx 0.5$  &	$18^2$ & $16^2$ & $13^2$ & $11^2$& $8^2$\\
		\hline 
	\end{tabular} 
	\caption{Required observation area given estimates of intensities $\lambda_1$ and $\lambda_2$, when expected interaction has strength $b=0.25$ and range $2\rsigma = 10$, testing with $K_{12}(r=7)$ at level $\alpha=5\%$ and requiring power at least 90\%.}
	\label{tab:area}
\end{table}

\subsection{Additional factors}
\label{sec:other}
The geometry of the area has an effect on the estimator's variance and hence the power of the test, but according to the analytical formula the effect is relatively small. For example, if we change from a square shape to an elongated rectangle shape with equal area but width-to-height -ratio 3, and consider interaction $b=0.25$ and type I error level $\alpha=5\%$, the power drops from 33.2\% to 32.9\% with $2\rsigma=10$ and 78.0\% to 76.8\% with $2\rsigma=20$ for sample size $(80,80)$, and from 15.4\% to 15.3\% with $2\rsigma=10$ and 41.4\% to 40.5\% with $2\rsigma=20$ for sample size $(30,80)$.

Increasing the type I error level $\alpha$ increases the power as illustrated in Table \ref{tab:alpha}. From the table we can see that a 5\% increase in $\alpha$ can reduce the type II error $\beta=1-power$ by more than 10\%. So in scenarios where we can tolerate some extra false positive discoveries with the simultaneous decrease in false negatives, for example when pre-screening a large data set for more involved downstream analysis on found interacting pairs, adjustments to $\alpha$ should be considered. 

\begin{table}[hb]
	\centering
	\begin{tabular}{|r|r|r|r|r|r|r|}
		\hline
		\multicolumn{2}{|r|}{($n_1,n_2$) = (10,10)} & (30,30) & (50,50) & (80,80) & (30,50) & (30,80) \\ 
		\hline
		$\alpha=$1\% & 0.01 & 0.08 & 0.26 & 0.68 & 0.14 & 0.24 \\ 
		$\alpha=$5\% & 0.06 & 0.21 & 0.48 & 0.86 & 0.32 & 0.47 \\ 
		$\alpha=$10\% & 0.10 & 0.31 & 0.61 & 0.92 & 0.44 & 0.59 \\ 
		\hline
	\end{tabular}
	\caption{Power of the $K_{12}$ independence test at most powerful range for typical type I error levels $\alpha$. Interaction $b=0.5$, $2\rsigma = 10$.}
	\label{tab:alpha}
\end{table}

\bibliographystyleApp{plainnat}
\bibliographyApp{biv-power}

\end{document}